\documentclass[%
 reprint,
superscriptaddress,
 amsmath,amssymb,
 aps,
 showkeys,
]{revtex4-2}

\usepackage{graphicx}
\usepackage{dcolumn}
\usepackage{bm}
\usepackage{bbm}
\usepackage{xcolor}
\usepackage{lmodern}
\usepackage{upgreek}
\usepackage[T1]{fontenc}
\usepackage[normalem]{ulem}
\usepackage{amssymb}
\usepackage{mathrsfs}
\usepackage{tikz}
\usetikzlibrary{positioning, shapes.geometric}
\usepackage{scalerel,stackengine}
\usepackage{lipsum}
\usepackage{siunitx}
\usepackage{lineno}
\begin{document}

\preprint{APS/123-QED}

\title{Programmable nonlinear rotation of acoustically levitated chiral stars}

\author{Sue Shi}
\email{sue.shi@ist.ac.at}
\affiliation{
 Institute of Science and Technology Austria (ISTA), Am Campus 1, 3400 Klosterneuburg, Austria
}
\author{Maximilian C. H\"ubl}
\email{maximilian.huebl@ist.ac.at}
\affiliation{
 Institute of Science and Technology Austria (ISTA), Am Campus 1, 3400 Klosterneuburg, Austria
}
\author{Aarnav Panda}
\email{aarnav.panda@ist.ac.at}
\affiliation{
 Institute of Science and Technology Austria (ISTA), Am Campus 1, 3400 Klosterneuburg, Austria
}

\author{Thorfinn Barton}
\affiliation{
 Institute of Science and Technology Austria (ISTA), Am Campus 1, 3400 Klosterneuburg, Austria
}
\email{thorfinn.barton@ist.ac.at}

\author{Carl P. Goodrich}
\email{carl.goodrich@ist.ac.at}
\affiliation{
 Institute of Science and Technology Austria (ISTA), Am Campus 1, 3400 Klosterneuburg, Austria
}
\author{Scott R. Waitukaitis}%
\email{scott.waitukaitis@ist.ac.at}
\affiliation{
 Institute of Science and Technology Austria (ISTA), Am Campus 1, 3400 Klosterneuburg, Austria
}

\date{\today}

\begin{abstract}

Acoustic levitation enables objects to be suspended and manipulated without physical contact, providing a versatile platform for studying the dynamics of freely suspended matter. Yet levitated objects, from droplets and thin films to particle assemblies, frequently rotate spontaneously, and this axial rotation often remains difficult to predict or control. There is no general understanding of how the shape of a levitated object determines whether it rotates, in which direction, or how its rotation changes with acoustic amplitude. Here, we establish how particle geometry and acoustic amplitude govern spontaneous rotation of acoustically levitated objects. We work with a family of shapes we call ``chiral stars'', within which we systematically vary handedness and number of arms. In a symmetric acoustic field, stars that preserve in-plane mirror symmetry do not rotate, while those that break it spontaneously rotate with a direction set by handedness. Beyond this symmetry requirement, we demonstrate that geometry, as encoded by the number of star arms, couples with acoustic amplitude to produce complex, nonlinear dynamics, including spontaneous reversal of rotation direction. To explain these behaviors, we develop a minimal torque model constrained by independent measurements of rotational drag, which captures the observed responses across all star geometries. We then demonstrate predictive power by designing a ``hybrid star'' with a prescribed amplitude-dependent response and reversal voltage. Our results not only establish how particle geometry and acoustic amplitude combine to govern rotation in acoustically levitated systems, but also reveal unanticipated higher-order couplings with the acoustic field. Beyond explaining spontaneous rotation, they open a route toward programmable rotational dynamics in acoustically levitated matter.

\end{abstract}

\maketitle

Modern acoustic levitation emerged in the 1970s as a tool for contact-free manipulation of small objects \cite{whymark1975acoustic, wang1984containerless,trinh1985compact}. Since then, it has developed into a versatile experimental platform for studying freely suspended matter \cite{andrade2018review,andrade2020acoustic}. In its simplest form, an ultrasonic standing wave generates radiation forces that trap micron- to millimeter-sized objects at well-defined positions, with minimal constraints on material properties. By freeing objects from the need for physical support and confinement, acoustic levitation has opened a window into an otherwise inaccessible array of exotic physical phenomena, ranging from nonreciprocal interactions and emergent dynamics to time crystals \cite{lim2019cluster,lim2024acoustic, wu2025nonreciprocity, king2025scattered,shi2025electrostatics,brown2026nonreciprocal,morrell2026nonreciprocal,gronert2026emergent}. At the same time, acoustic levitation is developing into a potentially transformative tool for applications in chemical and pharmaceutical analysis \cite{santesson2004airborne, wood2005portable,benmore2011amorphization, van2021quantitative} and the assembly of microscale structures \cite{vandaele2005non,caleap2014acoustically}. With recent advances in acoustic holography \cite{melde2016holograms} and real-time sound-sculpting \cite{marzo2015holographic}, these capabilities promise to expand even further, enabling increasingly sophisticated control over levitated matter. \par

Yet even as our sophistication advances, the dynamics of levitated objects can be surprisingly unpredictable, particularly with regard to spontaneous rotation, which has been observed across a wide range of levitated systems \cite{biswas1991rotation,baer2011analysis,kepa2022acoustic,lim2022mechanical,shi2025electrostatics}. The torques underlying such rotation have been attributed to oscillational instabilities \cite{rudnick1990oscillational}, acoustic-field asymmetry \cite{xiu2017rotation}, acoustic streaming \cite{trinh1994experimental}, and scattering-mediated interactions \cite{king2025scattered}. However, these explanations are largely system-specific and offer no general understanding of when an object can be expected to rotate, in what direction, and how fast.   Without this understanding, spontaneous rotation remains both difficult to suppress when it is a nuisance and to control when it is desired \cite{marzo2015holographic, lamprecht2015viscous,  hong2017dynamics, cox2018acoustic}. \par

In this work, we establish how particle geometry and acoustic amplitude govern spontaneous rotation in an air-based single-axis levitator. We work with a family of star-shaped particles with curved arms, which we call ``chiral stars'', allowing us to tune handedness and field coupling cleanly through the number of arms. We show that, in a symmetric acoustic field, broken in-plane mirror symmetry determines whether and in which direction a star rotates, while arm number controls the nonlinear response to acoustic amplitude, including unanticipated reversals of rotation direction. A minimal torque model constrained by independent measurements of rotational drag captures these behaviors and enables us to design a ``hybrid star'' with a prescribed reversal response, demonstrating predictive control of rotational dynamics through particle geometry.

\begin{figure*}[t!]
\centering
\includegraphics{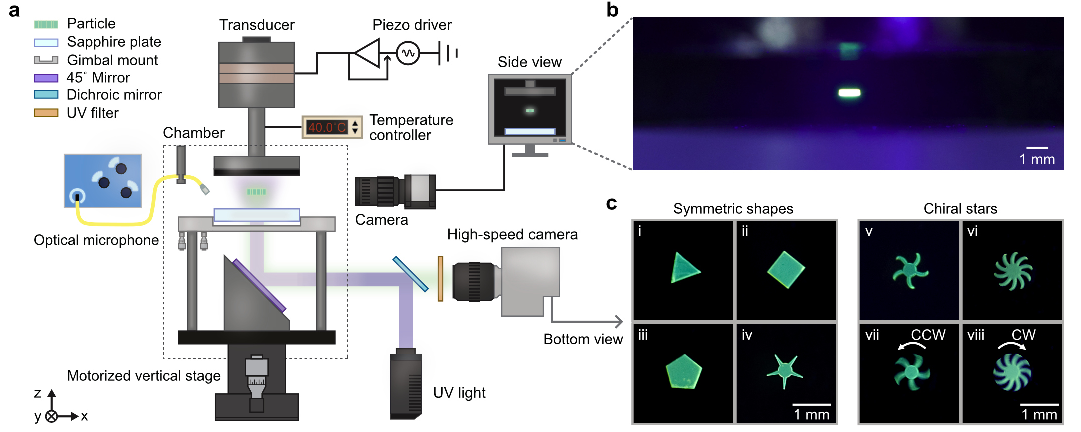}
\caption{\textbf{Acoustic levitation and rotation of 3D-printed particles.} \textbf{a,} Schematic of the experimental setup. A 3D-printed particle levitates at the nodal plane of a 51~kHz standing wave, which is generated between a Langevin transducer and a sapphire plate reflector. The transducer neck is held at 40~$\pm~2^\circ$C by a proportional–integral–derivative (PID) temperature controller. The reflector is secured to a gimbal mount and placed on a motorized vertical translation stage for precise alignment of the acoustic cavity. The sensor head of an optical microphone is placed immediately outside the cavity to monitor acoustic pressure. Ultraviolet (UV) illumination excites fluorescent 3D-printed particles, which are imaged from the side with a USB camera and from below with a high-speed camera via a $45^{\circ}$ mirror. \textbf{b,} Side-view image of a UV-illuminated fluorescent particle levitating at the nodal plane between the transducer horn and reflecting sapphire plate. \textbf{c,} Bottom-view images of symmetric shapes (i--iv) and chiral stars (v--viii). Mirror-symmetric particles, including triangles, squares, pentagons, and achiral stars, do not rotate spontaneously in an aligned cavity. Chiral stars (v–viii) rotate spontaneously immediately upon levitation, with the direction set by the handedness of the curved arms. Scale bars: 1~mm.}
\label{fig1}
\end{figure*}

The experimental setup consists of a single-axis acoustic levitation cavity with optical access for side- and bottom-view imaging, as shown in Fig.~\ref{fig1}a. We operate a Langevin-type transducer near its electromechanical resonance to generate a 51~kHz standing wave between the transducer horn and a reflecting sapphire plate. Particles levitate near the pressure node, approximately halfway between the horn and reflector. Crucially, the reflector is affixed to a custom-designed, precision gimbal mount for ultra-fine cavity alignment. The gimbal mount and reflector are furthermore on a motorized vertical stage for fine adjustment of the cavity height. An optical microphone is positioned immediately outside the acoustic cavity to monitor the acoustic pressure while minimizing perturbations to the field. A USB camera records the side view of the levitated particle (Fig.~\ref{fig1}b), while a high-speed camera records the bottom view via a 45$^{\circ}$ mirror. Representative bottom-view images are shown in Fig.~\ref{fig1}c. All particles are 200~\textmu m thick and are designed to fit within a 1~mm-diameter circle.\par

We first establish that spontaneous rotation requires broken symmetry in either the particle geometry or the acoustic field. To do so, we begin with mirror-symmetric particles and use the gimbal-mounted reflector to carefully align the acoustic cavity. Under optimal alignment, none of the symmetric polygons---triangles, squares, and pentagons---rotate spontaneously at any tested voltage (Supplementary Videos 1--3). Adding arms while preserving in-plane mirror symmetry also produces no rotation, as demonstrated by the achiral star in Fig.~\ref{fig1}c, panel iv (Supplementary Video 4). Tellingly, we can induce rotation in any of these shapes by intentionally misaligning the cavity, demonstrating that asymmetry in the acoustic field itself can generate rotational torques (Supplementary Video 5). By contrast, chiral stars rotate spontaneously even in an optimally aligned cavity. Within a few hundred milliseconds, they reach a steady rotation frequency, with the direction determined by the handedness of their curved arms (Fig.~\ref{fig1}c, panels vii and viii; Supplementary Videos 6 and 7). Together, these observations demonstrate that spontaneous rotation is not an inevitable consequence of acoustic levitation generally, but instead arises from broken symmetry in either the particle or the acoustic field. They further suggest that spontaneous rotation observed in many systems may be suppressed through more sophisticated control of the acoustic field.
\par 

\begin{figure*}[t!]
\centering
\includegraphics{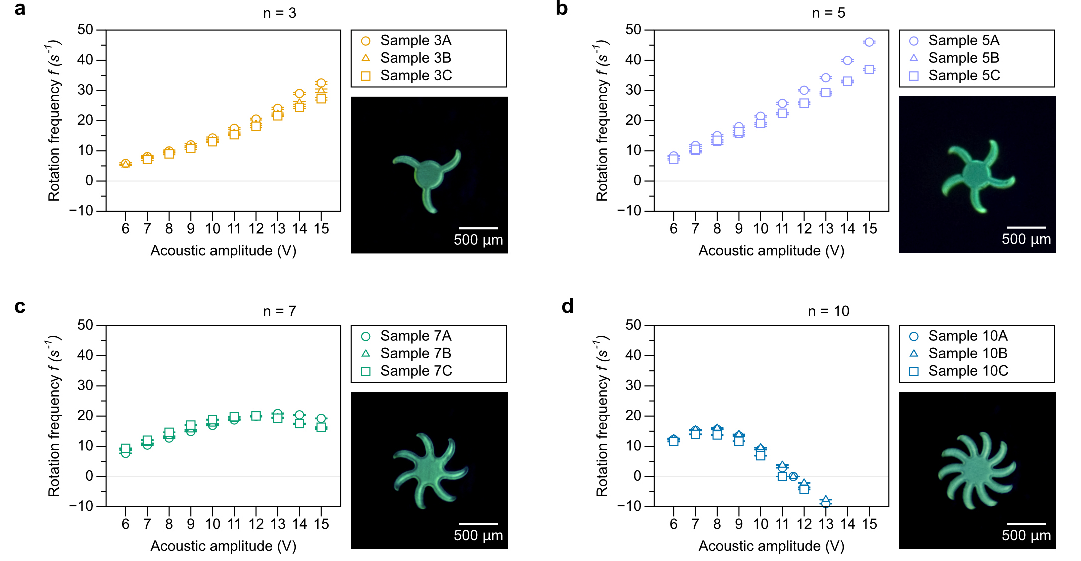}
\caption{\textbf{Geometry- and amplitude-dependent rotation of acoustically levitated chiral stars.} Steady-state rotation frequency is measured in revolutions per second ($s^{-1}$) as a function of the acoustic amplitude ($V$), for 3-armed (\textbf{a}), 5-armed (\textbf{b}), 7-armed (\textbf{c}), and 10-armed (\textbf{d}) chiral stars. Positive rotation is defined as counterclockwise (CCW) when viewed from below. Each panel shows measurements from three nominally identical 3D-printed particles. Error bars indicate the propagated standard error of the mean rotation frequency at each voltage, obtained by combining the per-video standard errors from frame-to-frame angular-velocity measurements. Representative bottom-view fluorescence images are shown to the right of each plot. Scale bars, 500~\textmu m.}
\label{fig2}
\end{figure*}

Having established broken mirror symmetry as the requirement for spontaneous rotation in a symmetric acoustic field, we next examine how particle geometry and acoustic amplitude together govern the rotational response. We fabricate a family of chiral stars with the same outer diameter and thickness but different numbers of identical arms. For consistency, we orient all particles to rotate counterclockwise when viewed from below (Fig.~\ref{fig2}, insets); flipping a star reverses its direction without changing its amplitude dependence (Fig.~S2). We measure the steady-state rotation frequency as a function of the voltage applied to the transducer, $V$, which is proportional to pressure (see Supplementary Information for tracking algorithm and Fig.~S3 for pressure measurements). For each geometry, measurements are repeated on three nominally identical particles to assess sample-to-sample variability arising from small fabrication differences. \par

Intuitive arguments might suggest that increasing the number of arms increases the surface area available to couple to the acoustic field and therefore should increase the rotation frequency at a given acoustic amplitude. Likewise, for a fixed geometry, one might naively expect that increasing acoustic amplitude should increase the driving torque, producing a monotonic increase in rotation frequency. Our measurements show that neither of these expectations holds in general. Stars with 3 and 5 arms rotate progressively faster over our accessible voltage range~(Fig.~\ref{fig2}a, b). By contrast, 7-armed stars reach a maximum rotation frequency near 12--13~$\mathrm{V}$ and slow at higher amplitude~(Fig.~\ref{fig2}c). For 10-armed stars, the turnover occurs at a much lower amplitude, near 7--8~$\mathrm{V}$, after which the rotation frequency decreases, crosses zero, and even reverses direction~(Fig.~\ref{fig2}d). Stars with 12 arms follow the same sequence at even lower amplitude, turning over at 7~$\mathrm{V}$ and reversing near 10~$\mathrm{V}$ (Fig.~S4). Adding arms therefore does not simply increase the driving torque, nor does increasing the acoustic amplitude necessarily increase the rotation frequency. \par

\begin{figure*}[t!]
\centering
\includegraphics{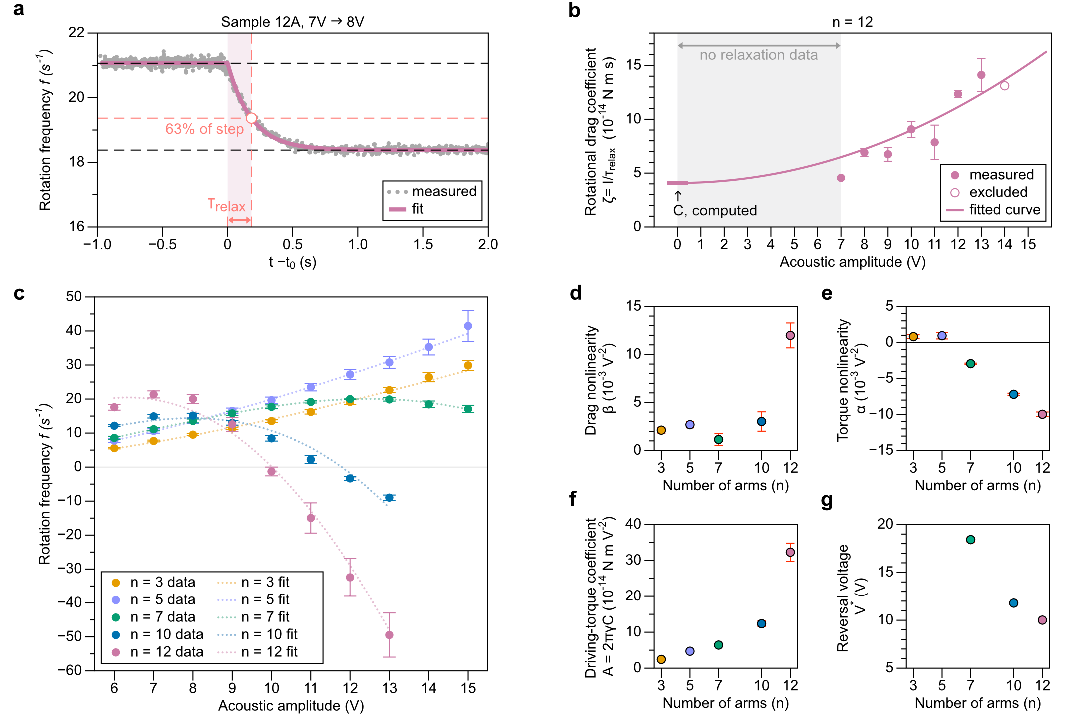}
\caption{\textbf{Rotational drag and geometry-dependent acoustic torques.} \textbf{a,} Representative relaxation measurement for a 12-armed star following a step in acoustic amplitude from 7 to 8~$\mathrm{V}$. Gray dotted line shows the measured rotation frequency and the solid pink curve shows the exponential fit. Black horizontal dashed lines indicate the initial and final steady-state rotation frequencies; the relaxation time $\tau_\mathrm{relax}\approx183~\mathrm{ms}$ is defined by the time required to reach 63\% of the step. \textbf{b,} Effective rotational drag coefficient, $\zeta = I/\tau_\mathrm{relax}$, as a function of acoustic amplitude for the 12-armed stars. Circles show the mean over three nominally identical stars. Filled symbols are included in the fit; open symbols mark voltages at which only one star yielded a valid relaxation measurement and are excluded. The solid curve shows the fit according to equation \eqref{zetaV}, with the zero-amplitude limit coefficient C computed via a boundary-element Stokes-flow method. The shaded region indicates the voltage range with no relaxation data. \textbf{c,} Rotation frequency as a function of acoustic amplitude for stars with $\mathrm{n} =$ 3, 5, 7, 10, and 12 arms, all averaged over three nominally identical stars. Points show the measured data and dotted curves show fits to the steady-state torque balance model given in equation \eqref{omegafit}. \textbf{d--f,} Geometry dependence of the fitted model parameters: drag nonlinearity $\beta = D/C$ (\textbf{d}), torque nonlinearity $\alpha = B/A$ (\textbf{e}), and leading acoustic torque coefficient $A = 2\pi\gamma C$ (\textbf{f}). \textbf{g,} Reversal voltage $V^* = 1/\sqrt{-\alpha}$ predicted from the fits for geometries with $\alpha < 0$. Error bars denote the standard error across three nominally identical stars (\textbf{b, c}) or the uncertainties of the fitted parameters (\textbf{d--g}).}
\label{fig3}
\end{figure*}

To model these complex behaviors, we begin with two qualitative observations. First, steady-state rotation implies a balance between driving torque and dissipation. In the simplest scenario, with constant driving torque $\tau_\mathrm{dr}$ and rotational drag coefficient $\zeta$, the dynamics would be described by
\begin{equation}\label{eq}
I\frac{df}{dt}= \frac{\tau_\mathrm{dr}}{2\pi} - \zeta f \,,
\end{equation}
where $I$ is the moment of inertia and $f$ is the rotation frequency. The steady-state rotation frequency is then simply 
\begin{equation}
f =\frac{\tau_\mathrm{dr}}{2\pi \zeta} \,.
\label{eq:torquebal}
\end{equation}

Second, both the driving torque and dissipation can in principle depend on particle geometry and acoustic amplitude. The reversals of the 10- and 12-armed stars imply that the driving torque can change sign at high acoustic amplitude, since drag can slow rotation but cannot reverse it. Moreover, oscillatory flows are known to alter the drag on immersed bodies in an amplitude-dependent way \cite{tunstall1968retardation}. To account for these effects, we expand both the driving torque and rotational drag coefficient in powers of acoustic pressure,~$p$, or equivalently~$V$. Because the time-averaged response is invariant under reversal of the oscillatory acoustic field, these expansions contain only even powers of $V$, \textit{i.e.},
\begin{align}
    \label{tauV}\tau_\mathrm{dr}(V) &= AV^2 + BV^4 + \ldots, \\
    \label{zetaV}\zeta(V) &= C + DV^2+ \ldots \,.
\end{align}
The driving torque has no constant since it vanishes in the absence of the acoustic field, whereas $C$ represents the finite rotational drag coefficient in the zero-amplitude limit.

To constrain these expansions, we experimentally characterize the rotational drag. For a given star, we measure how its rotation frequency relaxes following a step in acoustic amplitude. As shown for a representative 12-armed star in Fig.~\ref{fig3}a, following a voltage step from \qty{7}{\volt} to \qty{8}{\volt}, the rotation frequency relaxes exponentially toward a new steady-state value, with characteristic time $\tau_\mathrm{relax}$ ($\sim$\qty{183}{\milli\second} in Fig.~\ref{fig3}a). By contrast, optical microphone measurements show that the acoustic amplitude reaches $90\%$ of its new value within $\sim$\qty{11}{\milli\second} at each voltage step, much faster than the measured relaxation times (Fig.~S5). We therefore treat the driving torque and drag coefficient as constant during each relaxation, reducing the dynamics to equation~\eqref{eq} and allowing us to determine $\zeta$ directly from the measured relaxation (Fig.~S6). Extracting $\zeta$ across the accessible voltage range shows that it increases systematically with acoustic amplitude (Fig.~\ref{fig3}b), with similar behaviors observed for stars across all other arm numbers (Fig.~S7). These data indicate that the second-order expansion in equation~\eqref{zetaV} is sufficient to characterize the drag. By independently calculating the zero-amplitude coefficient $C$ from the particle geometry using a boundary-element Stokes-flow method, we extract the functional form of the drag for each particular geometry with the second-order term ($D$) as the only free parameter (see Supplementary Information).

With an understanding of the drag, we return to steady-state rotation curves to infer the driving torque responsible for the non-monotonic behavior. The simplest possible form for the driving-torque expansion is fourth order, \textit{i.e.}, $\tau_\mathrm{dr}(V) = AV^2 + BV^4$, since anything less cannot exhibit sign reversal. Substituting this expression for the torque and the established second-order expression for the dissipation into equation~\eqref{eq:torquebal}, the simplest predicted form for the steady-state rotation frequency as a function of acoustic amplitude is
\begin{equation}\label{omegafit}
    f(V) = \gamma V^2 \, \frac{1  + \alpha V^2}{1 + \beta V^2} \,,
\end{equation}
where $\gamma = A/(2\pi C)$, $\alpha = B/A$, and $\beta = D/C$. Because $C$ is computed from the geometry and $D$ is obtained from the drag measurements, $\beta$ is determined independently, leaving $\gamma$ and $\alpha$ as the only free parameters. To determine these, we fit equation~\eqref{omegafit} directly to the data from Figs.~\ref{fig2} and~S4, with the results shown in Fig.~\ref{fig3}c. The resulting fits capture the  distinct responses across all arm numbers, from the monotonic growth for stars with few arms to turnover and reversal for stars with many arms. As a robustness check, we also fit the rotation curves while allowing $\beta$ to vary as a third free parameter and find that the resulting curves are in excellent agreement with those obtained using the experimentally determined drag coefficients (see Fig.~S8). \par

The fitted parameters reveal how increasing the number of arms modifies the balance between driving torque and dissipation. First, rotational drag becomes increasingly sensitive to acoustic amplitude as the number of arms increases. This is captured by $\beta$, which remains relatively small for sparsely armed stars but increases substantially for more densely armed stars (Fig.~\ref{fig3}d). Second, the nonlinear contribution to the driving torque becomes increasingly negative with arm number. This behavior is captured by $\alpha=B/A$, which decreases from weakly positive values for the 3- and 5-armed stars to increasingly negative values for the 7-, 10-, and 12-armed stars (Fig.~\ref{fig3}e). For $\alpha<0$, this nonlinear contribution opposes the leading-order torque and can ultimately reverse its sign. Third, the leading-order driving torque increases strongly with arm number, explaining why densely armed stars rotate faster at low acoustic amplitude. This is reflected in the torque coefficient $A=2\pi\gamma C$, which increases systematically with arm number (Fig.~\ref{fig3}f). Thus, although densely armed stars experience stronger driving torque at low amplitudes, this initial advantage is eventually overcome by the increasingly negative nonlinear contribution. For $\alpha<0$, the driving torque vanishes when $1+\alpha V^2=0$, giving a reversal voltage
\begin{equation}
    V^*=\frac{1}{\sqrt{-\alpha}}\,.
\end{equation}
The inferred reversal voltage decreases with increasing arm number (Fig.~\ref{fig3}g), consistent with the earlier onset of turnover and reversal for more densely armed stars.\par

\begin{figure}[t!]
\centering
\includegraphics{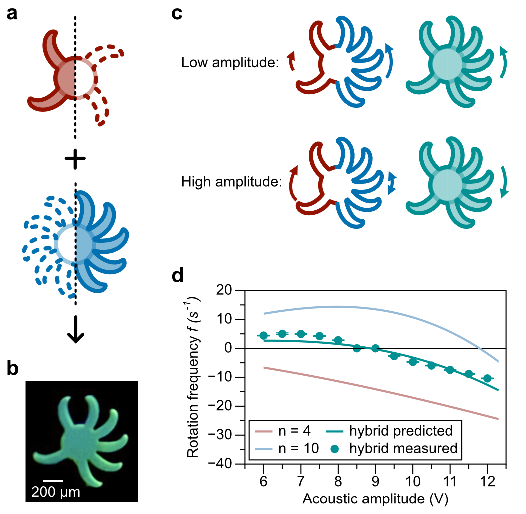}
\caption{\textbf{Programmable rotation behavior with a hybrid star.} \textbf{a,} Design principle of a hybrid star combining half of a 4-armed star and half of a 10-armed star with opposite handedness. \textbf{b,} Bottom-view fluorescent image of the fabricated hybrid star. Scale bar: 200~\textmu m. \textbf{c,} Predicted amplitude-dependent rotation response: at low acoustic amplitude, the half 10-armed side dominates due to stronger low-amplitude torque and the hybrid star rotates CCW; at high acoustic amplitude, the CCW contribution from the half 10-armed side either becomes strongly suppressed or fully reverses, and the hybrid star rotates CW. \textbf{d,} Rotation frequency as a function of acoustic amplitude. The red curve is the interpolated rotation response of CW 4-armed stars ($f_4^{\text{CW}}(V)$) and the blue curve is the fitted rotation response of CCW 10-armed stars ($f_{10}^{\text{CCW}}(V)$). The predicted response of the hybrid star in \textbf{b} (turquoise curve) is their equally weighted sum, equation~\eqref{hybrid}, and turquoise circles show the measured response. The hybrid star rotates CCW at low amplitude and CW at high amplitude, passing through zero at the predicted reversal voltage.}
\label{fig4}
\end{figure}

Having used our model to explain the observed rotation curves, we now ask whether it can also be used to design particles with prescribed rotational dynamics.  We test this idea by designing a ``hybrid star'' that combines two chiral motifs with competing amplitude-dependent responses. One half is taken from a 4-armed star with clockwise handedness, while the other is taken from a 10-armed star with counterclockwise handedness (Fig.~\ref{fig4}a). A bottom-view image of the resulting hybrid 7-armed star is shown in Fig.~\ref{fig4}b. \par

To predict the hybrid star's response, we combine the fitted rotation curves of its constituent geometries. The response of the clockwise 4-armed motif is estimated by interpolating between the fitted 3- and 5-armed stars and reversing the sign to account for its opposite handedness (Figs.~S9 and S10), while the counterclockwise 10-armed response is taken directly from Fig.~\ref{fig3}c. As a first approximation, we assume the two motifs contribute equally, giving
\begin{equation}\label{hybrid}
f_\mathrm{hybrid}(V) =
\frac{1}{2}f_4^\mathrm{CW}(V)
+ \frac{1}{2}f_{10}^\mathrm{CCW}(V)\,.
\end{equation}
This approximation predicts the rotation response for the hybrid star. At low amplitude, the 10-armed motif dominates, producing counterclockwise rotation. As the amplitude increases, its contribution weakens relative to that of the 4-armed motif, eventually producing a reversal to clockwise rotation (Fig.~\ref{fig4}c). Measurements with a fabricated hybrid star correspond remarkably well to this predicted behavior, including the reversal of rotation direction and excellent agreement between the predicted and measured reversal voltages (Fig.~\ref{fig4}d, Supplementary Video 8). This demonstrates that the geometry-dependent responses of individual stars can be combined to predict and design new complex dynamics.

Our experiments establish how particle geometry and acoustic amplitude govern spontaneous rotation in acoustic levitation. In a well-aligned single-axis standing-wave cavity, broken in-plane mirror symmetry is required for rotation, with handedness determining its direction. Conversely, misalignment can induce rotation even in mirror-symmetric particles, potentially explaining why spontaneous rotation is so widely observed. Beyond these symmetry considerations, we show that geometry also controls the nonlinear response to acoustic amplitude. For our chiral stars, increasing the number of arms strengthens the leading low-amplitude torque while introducing an increasingly strong opposing contribution at higher amplitudes, causing rotation to slow and ultimately reverse. By experimentally characterizing rotational drag, we disentangle the competing torque contributions and develop a model that captures all observed behaviors. We demonstrate the predictive power of this model by designing a hybrid star with a prescribed amplitude-dependent response.\par

The physical origin of the opposing high-amplitude torque remains unresolved. Importantly, the observed reversal cannot arise within the usual leading-order description of acoustic radiation and streaming. Standard treatments begin with a linear, first-order oscillatory acoustic field and retain the time-averaged response to second order in acoustic amplitude, while neglecting higher-order terms \cite{bruus2011acoustofluidics}. This perturbative approximation successfully describes acoustic radiation forces and torques in many weakly nonlinear systems, including cases where multiple scattering modifies the underlying acoustic field \cite{silva2014acoustic,mitri2017acoustic}, and likewise underlies standard descriptions of acoustic streaming \cite{lighthill1978acoustic,muller2015theoretical}. Within this description, superpositions of second-order responses cannot account for the observed amplitude-driven reversal: second-order contributions can reinforce or oppose one another, but if they share the same quadratic pressure dependence, their relative strengths remain fixed as the pressure is increased. The reversal therefore points to a higher-order correction, represented in our model by the $p^4$ contribution. The systematic strengthening of this higher-order contribution with arm number suggests that nonlinear effects become increasingly important as neighboring arms are brought closer together, potentially reflecting amplitude-dependent coupling between adjacent features of the particle. Resolving the microscopic origin of this coupling will require direct measurements or simulations of the acoustic and surrounding flow fields.\par

\section*{Methods}

The acoustic levitation cavity was created by a bolt-clamped Langevin transducer based on the design of Andrade \textit{et al.} \cite{andrade2020design}, with piezoelectric elements from a commercial transducer (STH-S44-3030-A, Micromechatronics). The transducer was operated with a feedback-controlled amplifier (PDUS210-FLEX, PiezoDrive) near its electromechanical resonance, $f_0 \approx 51~\mathrm{kHz}$, corresponding to an acoustic wavelength of $\lambda_0 \approx~$\qty{6.8}{\milli\metre}. The transducer horn was \qty{25.4}{\milli\metre} in diameter and \qty{3}{\milli\metre} thick, coated with black paint to improve optical contrast. The reflector was a sapphire window measuring \qty{25.4}{\milli\metre} in diameter and \qty{5}{\milli\metre} thick (WG30530, Thorlabs), mounted on a custom-modified gimbal optical mount (U200-G2K, Newport). To precisely control the vertical position of the reflector, the gimbal mount was further attached via an optical caging system to a motorized translation stage driven by a DC servo motor (Z925B, Thorlabs) and controller (KDC101, Thorlabs). The horn-reflector distance was adjusted to approximately $\lambda_0/2\approx$ \qty{3.4}{\milli\metre} and fine-tuned to cavity resonance using an optical microphone (Eta100 Ultra, Xarion Laser Acoustics). The microphone was also used to monitor the acoustic pressure during experiments, and was placed immediately outside the acoustic cavity to avoid disturbances to the acoustic field inside the cavity. Because the resonance frequency of the transducer was sensitive to changes in its temperature caused by varying the acoustic amplitude, we maintained the transducer temperature at $40\pm2^\circ$C with a silicone heating pad (50 $\times$ \qty{100}{\milli\metre}, \qty{12}{\volt}, \qty{15}{\watt}, Amazon), a type-K thermocouple (303-90-961, RS Components), and a custom-built PID temperature controller. Particles were illuminated using a collimated UV LED (M405LP1-C1, Thorlabs). The fluorescent light was isolated from the UV light source via a dichroic mirror (DMLP425L, Thorlabs) and a glass UV filter (FGL420S, Thorlabs). Side-view imaging was performed with a digital camera (DCU224M-GL, Thorlabs), while bottom-view dynamics were recorded through a $45^\circ$ mirror using a high-speed camera (Phantom V1612, Vision Research) at 3000 frames per second (fps). Experimental control and data acquisition were performed using a multifunction DAQ device (USB-6002, National Instruments). See Supplementary Information for particle fabrication details.\par

\section*{Acknowledgments}

We thank T. Baasch for insightful discussions. We acknowledge the MIBA machine shop at the Institute of Science and Technology Austria for instrumentation support.

\section*{Funding}

This research was funded in whole or in part by the Austrian Science Fund (FWF) PAT2433425. For Open Access purposes, the author has applied a CC BY public copyright license to any author accepted manuscript version arising from this submission.  M.C.H. and C.P.G. acknowledge funding by the Gesellschaft f\"ur Forschungsf\"orderung Nieder\"osterreich under project FTI23-G-011.

\section*{Author Contributions}
S.S. and S.R.W conceived the research. S.S. conducted the experiments. M.C.H., C.P.G, S.S., and S.R.W. developed the theoretical framework and contributed to the experimental design. A.P. and T.B fabricated all samples. S.S. analyzed the data. M.C.H. developed the particle rotation tracking algorithm. S.S., S.R.W, and M.C.H. wrote the manuscript. All authors discussed and provided input to the manuscript.

\end{document}

\preprint{}

\title{Supplementary Information: Programmable nonlinear rotation of acoustically levitated chiral stars}

\author{Sue Shi}
\email{sue.shi@ist.ac.at}
\affiliation{
 Institute of Science and Technology Austria (ISTA), Am Campus 1, 3400 Klosterneuburg, Austria
}
\author{Maximilian C. H\"ubl}
\email{maximilian.huebl@ist.ac.at}
\affiliation{
 Institute of Science and Technology Austria (ISTA), Am Campus 1, 3400 Klosterneuburg, Austria
}
\author{Aarnav Panda}
\email{aarnav.panda@ist.ac.at}
\affiliation{
 Institute of Science and Technology Austria (ISTA), Am Campus 1, 3400 Klosterneuburg, Austria
}

\author{Thorfinn Barton}
\affiliation{
 Institute of Science and Technology Austria (ISTA), Am Campus 1, 3400 Klosterneuburg, Austria
}
\email{thorfinn.barton@ist.ac.at}
\author{Carl P. Goodrich}
\email{carl.goodrich@ist.ac.at}
\affiliation{
 Institute of Science and Technology Austria (ISTA), Am Campus 1, 3400 Klosterneuburg, Austria
}
\author{Scott R. Waitukaitis}%
\email{scott.waitukaitis@ist.ac.at}
\affiliation{
 Institute of Science and Technology Austria (ISTA), Am Campus 1, 3400 Klosterneuburg, Austria
}

\maketitle

\section{Particle Fabrication}
Particles were fabricated by two-photon polymerization using a NanoOne printer (UpNano GmbH) and UpPhoto photopolymer resin. UpPhoto is fluorescent under UV illumination, providing high contrast for optical tracking. Particle geometries were designed in Autodesk Fusion and exported as high-resolution STL files. Both the symmetric control particles and the chiral stars were inscribed within a 1~mm-diameter circle and extruded to a thickness of 200~\textmu m. Symmetric particles have equal side lengths within each shape. For the chiral stars, individual arms were generated using a circular pattern to ensure equal angular spacing. Printing was performed using a Leica \(10\times\), NA 0.4 objective with a  \qty{780}{\nano\metre} laser. Particles were printed onto a glass substrate using the woodpile infill option. Following printing, the substrate was immersed in isopropyl alcohol for 10 min to remove unpolymerized resin. The particles were then released from the substrate and handled using fine tweezers.

\section{Particle rotation tracking}

\begin{figure}[h]
\centering
\includegraphics[width = 0.5\linewidth]{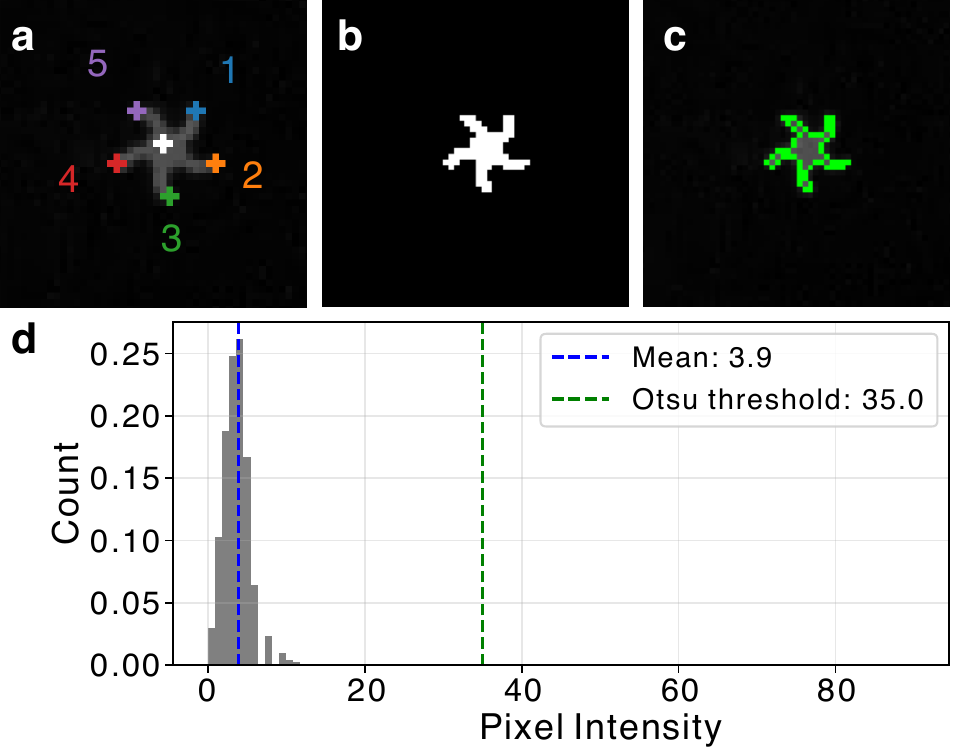}
\caption{\textbf{Image analysis for particle tracking.} The goal of this procedure is to obtain the positions of the center of mass (white) and the arm tips (colored), as shown in \textbf{a}. The points are extracted by first performing binarization on the image \textbf{b}, and then extracting the contour of the star (green pixels in \textbf{c}). Panel \textbf{d} shows the histogram of the image, and the resulting mean intensity (blue) and the binarization intensity found via Otsu's algorithm (green).}
\label{figS1}
\end{figure}

To measure the rotation frequency of the particles, we perform the following video analysis.
We first convert each frame to grayscale and binarize it using Otsu thresholding~\cite{otsu.1976} to separate the star from the background.
From the star’s outer contour we compute its centroid and, about it, the radial profile $r(\theta)$, i.e. the largest contour radius in each $0.25^\circ$ bin of polar angle, smoothed over $1.75^\circ$.
Since the number of arms $n$ is known, we divide the circle into $n$ equal angular sectors, positioned so that one is centered on the global maximum of $r(\theta)$, and, in each sector, interpret the contour points farthest from the centroid as the tip of a star arm.
The different stages of image analysis are shown in Fig.~\ref{figS1}.

The output of this procedure at frame $t$ is a list of arm tip positions $x^{(i)}_{t}$, where $i$ indexes the arms of the star.
The positions are given in the center-of-mass frame of the star and are ordered according to the angles $\theta^{(i)}_t$ they make with a fixed reference axis.
We use the notation $(i)$ to emphasize that this ordering is by angle rather than by arm identity: $x^{(1)}_{t}$ and $x^{(1)}_{t+1}$ do not necessarily correspond to the same physical arm.
To obtain a consistent ordering, we need a way to track arm identity between frames: we seek a mapping $i \to j$ that reorders the list of arm positions at time $t$ to match the ordering of the arms at time $t-1$.

A natural approach is to compute the angular displacement across a time step, $\Delta\theta^{(i, j)}_t = \theta^{(j)}_t - \theta^{(i)}_{t-1}$, and choose the mapping that assigns each arm at $t$ to the arm at $t-1$ with the smallest angular displacement.
However, due to limited image quality and time resolution, such a mapping does not always exist: in some frames, multiple arms at $t$ would be mapped to the same arm at $t-1$.
To resolve this, we consider all cyclic one-to-one mappings between the arms and, for each, count the number of ``satisfied arms'' -- arms that are mapped to their closest (in angular displacement) counterpart.
We then select the mapping with the highest number of satisfied arms to track arm identity across the frame pair.

Iterating this procedure yields a consistently ordered trajectory of arm tip positions $x^i_t$ and angles $\theta^i_t$, from which we extract the rotation frequency and other quantities of interest.

\section{Rotational Reynolds number}

The Reynolds number associated with star rotation was estimated using the particle's outer (tip) radius $R$ as the characteristic length and the tangential speed $U=\omega R$ as the characteristic velocity,
\begin{equation}
    \mathrm{Re} = \frac{\rho\,U R}{\mu} = \frac{\rho\,\omega R^2}{\mu} = \frac{\omega R^2}{\nu}\,,
\end{equation}
where $\rho$, $\mu$, and $\nu$ are the density, dynamic viscosity, and kinematic viscosity of air, respectively. At $27^\circ \mathrm{C}$, we take $\rho \approx 1.176\,\mathrm{kg\,m^{-3}}$ and $\mu \approx 1.85\times10^{-5}\,\mathrm{Pa\,s}$, giving $\nu \approx 1.57 \times 10^{-5}\,\mathrm{m^2\,s^{-1}}$. Writing $\omega = 2\pi f$, where $f$ is the rotation frequency in revolutions per second, and taking $R\approx 0.5\,\mathrm{mm}$, this yields
\begin{equation}
    \mathrm{Re} = \frac{2\pi |f| R^2}{\nu} \approx 0.1\,\mathrm{s} \, |f|\,.
\end{equation}
For the experimentally observed range $0\leq |f| \leq 50\, \mathrm{s^{-1}}$, the corresponding Reynolds number spans approximately $0\leq \mathrm{Re}\leq 5$, placing the rotational motion in the low-to-moderate Reynolds-number regime. For comparison, the leading inertial correction to the viscous torque on a rotating sphere of the same radius amounts to only $\sim 2\%$ at $\mathrm{Re}=5$~\cite{bickley1938lxv,collins1955steady, liu2010wall}, well below the scatter of our measurements. This suggests that inertial corrections to rotational drag may be modest over the experimentally accessible range. We therefore adopt a linear viscous drag law as a leading-order approximation, neglecting higher-order terms in $f$.

\section{Boundary-element Stokes-flow calculation}
We used the boundary-element method~\cite{Youngren_Acrivos_1975, Pozrikidis_1992} to calculate the drag coefficients $\zeta(0) = C$ in the Stokes limit. 
We formulated the Stokes problem as a single-layer boundary integral, with surfaces taken directly from the CAD geometry of the stars and the dynamic viscosity of air set to $\mu \approx 1.85\times10^{-5}\,\mathrm{Pa\,s}$.
The calculation was carried out with the Inti.jl library~\cite{inti_jl} and using GMRES~\cite{gmres} to solve the resulting linear systems.

\section{Drag fits}
At each step voltage, $\tau_\mathrm{relax}$ is averaged over three nominally identical stars of each geometry and converted to $\zeta = I/\tau_{relax}$. Voltages at which only one star yielded a valid relaxation measurement, either because the step was not recorded for the other stars or because their relaxation fits failed the quality criteria, are shown in Fig.~\ref{dragfits} but not included in the drag fits. Most rejected fits occur near the turnover ($\mathrm{n = 7}$) or the reversal ($\mathrm{n = 10}$), where the change in rotation frequency across a step is small or the star starts from rest, so the step time and relaxation time cannot be reliably separated. The drag data are fitted with equation~(4) of the main text by unweighted least squares, with $C$ fixed at the drag computed for each geometry in still air. For $\mathrm{n=7}$, $\zeta$ increases by less than a factor of two over the measured range and does not vary monotonically with voltage. $D$ is therefore only weakly constrained and the fitted curve should be interpreted as an upper bound on the drag nonlinearity.  \par

\newpage

\begin{figure*}[t!]
\centering
\includegraphics{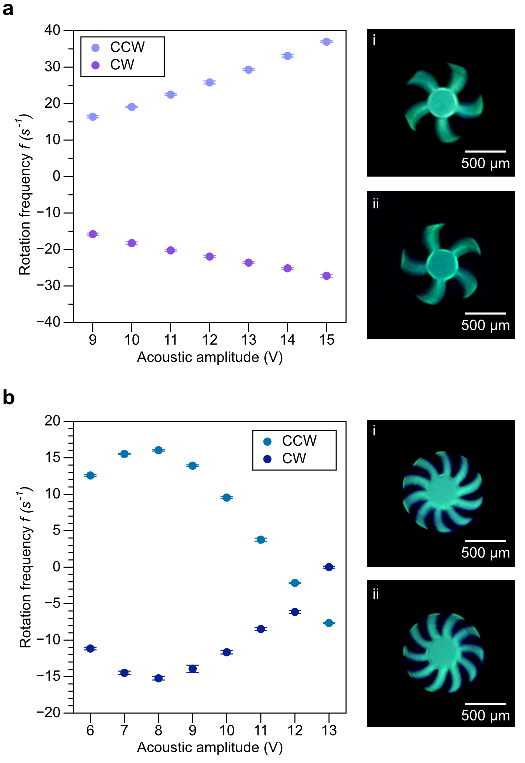}
\caption{\textbf{Flipping a chiral star reverses its rotation direction while preserving its amplitude-dependent response.} \textbf{a,} Rotation frequency ($s^{-1}$) as a function of acoustic amplitude~($\mathrm{V}$), for a 5-armed chiral star measured in two opposite orientations, corresponding to counterclockwise (CCW, pale purple) and clockwise (CW, bright purple) rotation when viewed from below. The orientations exhibit nearly equal rotation magnitudes with opposite signs over the measured voltage range. \textbf{b,} Corresponding measurements for a 10-armed star. Flipping reverses the sign of the rotation while preserving the characteristic non-monotonic amplitude dependence, including the turnover at $\sim$ \qty{8}{\volt}. Small differences between the two orientations likely reflect residual asymmetries in the acoustic field and particle loading. Representative bottom-view images show the CCW (i) and CW (ii) orientations. Error bars indicate the propagated standard error of the mean rotation frequency. Scale bars, 500~\textmu m.}
\label{flip}
\end{figure*}

\begin{figure*}[t!]
\centering
\includegraphics{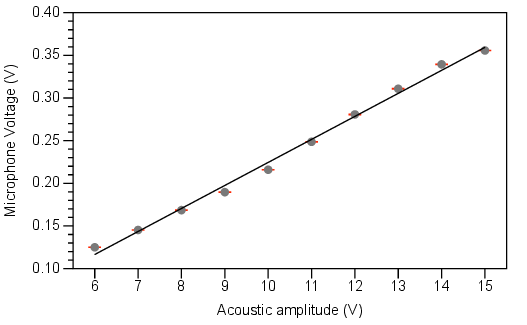}
\caption{\textbf{Acoustic pressure increases approximately linearly with acoustic amplitude.} Acoustic pressure is measured with the optical microphone immediately outside an empty cavity, because any levitated object perturbs the acoustic field \cite{leung1982resonance, curzon1986using, rudnick1990oscillational}. The microphone output voltage, which is proportional to acoustic pressure, is plotted against acoustic amplitude, defined as the voltage applied to the transducer, $\mathrm{V}$. The solid line is a linear fit over 6--15~$\mathrm{V}$. Error bars denote the standard deviation of 1000 readings at each voltage.}
\label{micmeas}
\end{figure*}

\begin{figure*}[t!]
\centering
\includegraphics{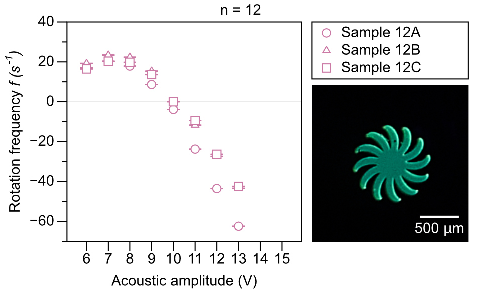}
\caption{\textbf{Rotation frequency of 12-armed chiral stars.} Rotation frequency as a function of acoustic amplitude for three nominally identical 12-armed stars. Positive rotation is defined as counterclockwise when viewed from below. The rotation frequency initially increases with acoustic amplitude, reaches a peak around 7~$\mathrm{V}$, and then decreases, crossing zero and reversing direction at higher amplitude. Error bars as in Fig.~2. A representative bottom-view fluorescence image is shown to the right. Scale bar, 500~\textmu m.}
\label{12arm}
\end{figure*}

\begin{figure*}[t!]
\centering
\includegraphics{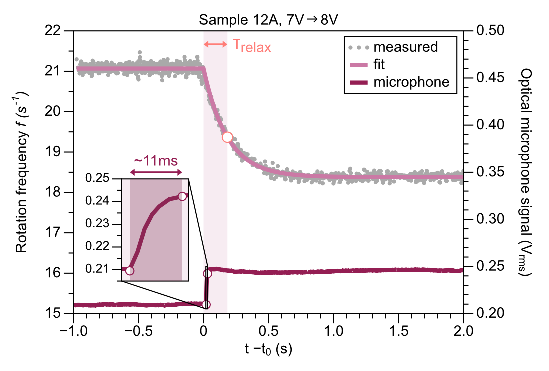}
\caption{\textbf{Response of the acoustic field compared with the star’s relaxation.} Rotation frequency of a 12-armed star after the acoustic amplitude is stepped from 7 to 8~$\mathrm{V}$ (gray points, left axis), with the exponential fit (pink line, left axis). The shaded band marks the relaxation time $\tau_\mathrm{relax} \approx$ 183~ms, defined as the time to reach 63\% of the change in rotation frequency (open circle). The dark line (right axis) is the root-mean-square (RMS) signal of the optical microphone for the same voltage step. Inset: the acoustic amplitude reaches 90\% of its new value ~11 ms after the step, an order of magnitude faster than $\tau_\mathrm{relax}$. Here 0\% and 100\% are the average signals over the first and last 1 s of the recording.}
\label{12arm_mic}
\end{figure*}

\begin{figure*}[t!]
\centering
\includegraphics{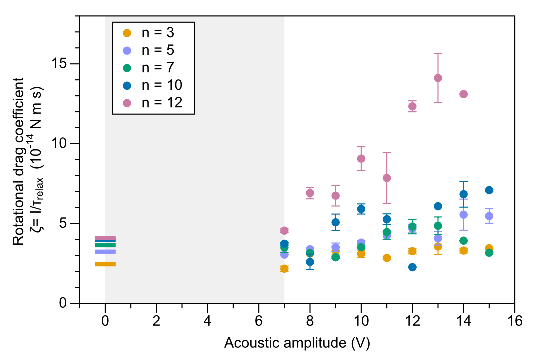}
\caption{\textbf{Rotational drag coefficient of all chiral star geometries as a function of acoustic amplitude.} The drag coefficient $\zeta = I/\tau_{relax}$ is obtained from the moment of inertia $I$ calculated for each star geometry and the relaxation time $\tau_\mathrm{relax}$ averaged over three nominally identical stars. Each point is plotted at the final voltage of the corresponding voltage step. Error bars denote the standard error across stars, and points without error bars contain data from a single star. At \qty{14}{\volt}, 12-armed stars become unstable and oscillate along the vertical axis, so no data are shown at \qty{15}{\volt}. No relaxation data exist in the shaded region, which marks the voltage range below that required for stable levitation. Dashes at \qty{0}{\volt} mark the zero-amplitude drag coefficient C computed for each star geometry with a boundary-element Stokes-flow method.}
\label{dragdata}
\end{figure*}

\begin{figure*}[t!]
\centering
\includegraphics{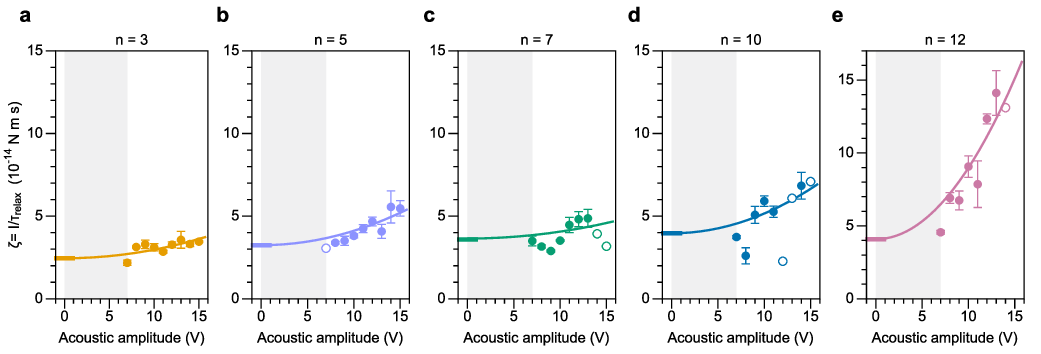}
\caption{\textbf{Acoustic amplitude dependence of the rotational drag coefficient.} Effective drag coefficient $\zeta$ as a function of acoustic amplitude $\mathrm{V}$ for stars with \textbf{a,} $\mathrm{n = 3}$, \textbf{b,} 5, \textbf{c,} 7, \textbf{d,} 10, and \textbf{e,} 12 arms. Filled circles show $\zeta$ computed from the mean relaxation time over three nominally identical stars, with error bars denoting the standard error across stars. Open circles mark voltages at which only one of the three stars yielded a valid relaxation measurement, either because the step was not recorded for the other stars due to levitation onset voltage differences or because their fits failed the quality criteria. These points are shown but not fitted. Solid curves are fits of $\zeta(V) = C + DV^2$ with zero-amplitude limit $C$ fixed at the value computed for still air (horizontal bar at $V = 0$). The shaded region marks the voltage range below the minimum required for stable levitation, and hence no relaxation data are available there.}
\label{dragfits}
\end{figure*}

\begin{figure*}[t!]
\centering
\includegraphics{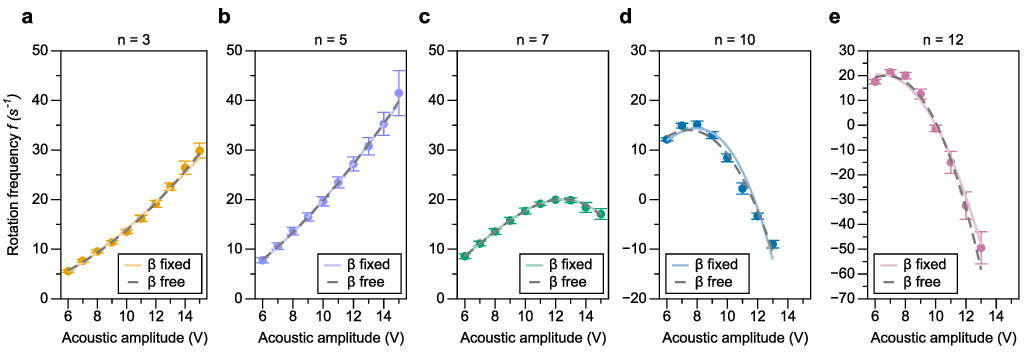}
\caption{\textbf{Rotation frequency fits with $\beta$ fixed and with $\beta$ free.} Rotation frequency as a function of acoustic amplitude for \textbf{a,} $\mathrm{n=3}$, \textbf{b,} 5, \textbf{c,} 7, \textbf{d,} 10, and \textbf{e,} 12 arms, averaged over three nominally identical stars (error bars as in Fig.~3c). Solid curves are fits of equation (5) with $\beta = D/C$ fixed by drag measurements; dashed curves leave $\beta$ as an additional free parameter.}
\label{betafree}
\end{figure*}

\begin{figure*}[t!]
\centering
\includegraphics{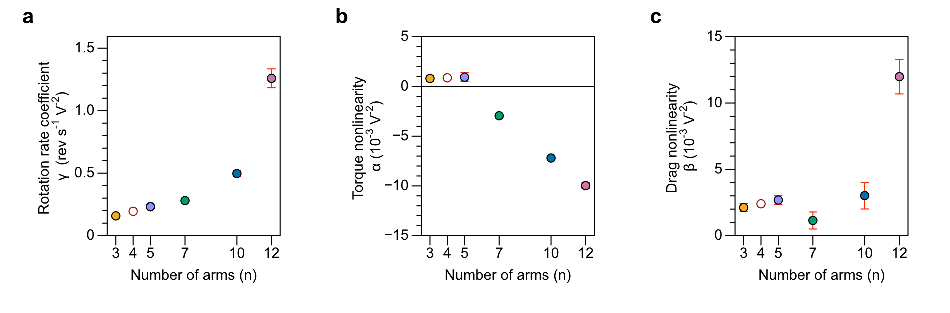}
\caption{\textbf{Interpolated model parameters for 4-armed stars.} Rotation-rate coefficient $\gamma$ (\textbf{a}), torque nonlinearity $\alpha$ (\textbf{b}), and drag nonlinearity $\beta$ (\textbf{c}) as a function of the number of arms, $\mathrm{n}$. Solid circles show $\gamma$ and $\alpha$ from fits of equation (5) to the measured rotation curves, and $\beta = D/C$ from the independent drag measurements. Open circles show the values for $\mathrm{n=4}$, obtained by linear interpolation between the 3- and 5-armed stars, which is the mean of their values. Interpolation is justified because all three parameters vary only modestly between $\mathrm{n=3}$ and $\mathrm{n=5}$ compared with the change at $\mathrm{n\geq 7}$. Error bars are as in Fig.~3.}
\label{4armpars}
\end{figure*}

\begin{figure*}[t!]
\centering
\includegraphics{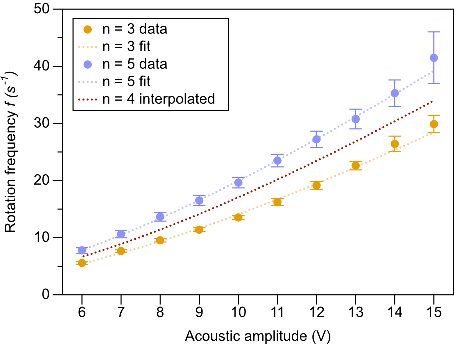}
\caption{\textbf{Predicted rotation curve of 4-armed stars.} Rotation frequency as a function of acoustic amplitude for 3- and 5-armed stars, as plotted in Fig.~3c. The dark red dotted line is the predicted curve for 4-armed stars from equation (5), using the interpolated parameters in Fig.~\ref{4armpars}. All curves show counterclockwise rotation; the clockwise 4-armed contribution used in equation (7) of the main text has the same magnitude but opposite sign.}
\label{4armfit}
\end{figure*}

\clearpage

\section*{Supplementary Videos}

\noindent \textbf{Supplementary Video 1:} Bottom view of the triangle shown in Fig.~1\textbf{c}, panel i, levitating without rotation in the aligned cavity at an acoustic amplitude of \qty{14}{\volt}. The video was recorded at 3000 frames per second and is played back in real time.
\bigskip

\noindent \textbf{Supplementary Video 2:} Bottom view of the square shown in Fig.~1\textbf{c}, panel ii, levitating without rotation in the aligned cavity at an acoustic amplitude of \qty{14}{\volt}. The video was recorded at 3000 frames per second and is played back in real time.
\bigskip

\noindent \textbf{Supplementary Video 3:} Bottom view of the pentagon shown in Fig.~1\textbf{c}, panel iii, levitating without rotation in the aligned cavity at an acoustic amplitude of \qty{14}{\volt}. The video was recorded at 3000 frames per second and is played back in real time.
\bigskip

\noindent \textbf{Supplementary Video 4:} Bottom view of the achiral star shown in Fig.~1\textbf{c}, panel iv, levitating without rotation in the aligned cavity at an acoustic amplitude of \qty{14}{\volt}. The video was recorded at 3000 frames per second and is played back in real time.
\bigskip

\noindent \textbf{Supplementary Video 5:} Bottom view of the square shape in Fig.~1c, panel ii rotating spontaneously in a misaligned cavity. Acoustic amplitude is kept at \qty{14}{\volt}. The video was recorded at 3000 frames per second and is played back in real time.
\bigskip

\noindent \textbf{Supplementary Video 6:} Bottom view of the 5-armed star shown in Fig.~S2\textbf{a}, switching from counterclockwise to clockwise rotation after being flipped. The acoustic amplitude is held constant at \qty{10}{\volt}. The video was recorded at 3000 frames per second and is played back at $0.01\times$ real time.
\bigskip

\noindent \textbf{Supplementary Video 7:} Bottom view of the 10-armed star shown in Fig.~S2\textbf{b}, switching from counterclockwise to clockwise rotation after being flipped. The acoustic amplitude is held constant at \qty{10}{\volt}. The video was recorded at 3000 frames per second and is played back at $0.01\times$ real time.
\bigskip

\noindent \textbf{Supplementary Video 8:} Bottom view of the hybrid star shown in Fig.~4\textbf{b}. At \qty{7}{\volt}, it rotates counterclockwise, following the curvature of the densely spaced arms. At \qty{9}{\volt}, it remains stationary. At \qty{11}{\volt}, it rotates clockwise, following the curvature of the sparsely spaced arms. The video was recorded at 3000 frames per second and is played back at $0.01\times$ real time during non-zero rotation and in real time when the particle is stationary.